\documentclass[preprint,showpacs,preprintnumbers,amsmath,amssymb]{revtex4}

\usepackage{graphicx}
\usepackage{dcolumn}
\usepackage{bm}

\begin{document}

\title{Absence of molecular mobility on nano--second time scales
in amorphous ice phases}
\author{
        M.M.~Koza$^{1}$,
        B.~Geil$^{2}$,
        H.~Schober$^{1}$,
        F.~Natali$^{1}$.
   }
\address{
         $^1$Institut Laue-Langevin, F-38042 Grenoble Cedex, France.\\
         $^2$Fachbereich Physik, TU Darmstadt, D-64289 Darmstadt, Germany.\\
                 }

\date{\today}

\begin{abstract}
High--resolution neutron backscattering techniques
are exploited to study the elastic and quasi-elastic response of
the high--density amorphous (HDA), the low--density amorphous (LDA)
and the crystalline ice I$_{\rm c}$ upon temperature changes.
Within the temperature ranges of their structural stability
(HDA at $T\le 80$~K, LDA at $T\le 135$~K, ice I$_{\rm c}$ at
$T < 200$~K) the Debye--Waller factors and mean--square displacements
characterise all states as harmonic solids.
During the transformations HDA$\rightarrow$LDA ($T\approx 100$~K),
LDA$\rightarrow$I$_{\rm c}$ ($T\approx150$~K) and the supposed
glass transition with $T_{\rm g}\approx 135$~K no relaxation
processes can be detected on a time scale $t<4$~ns.
It can be concluded from coherent scattering measurements (D$_2$O)
that LDA starts to recrystallise into ice I$_{\rm c}$ at
$T\approx135$~K, i.e. at the supposed $T_{\rm g}$.
In the framework of the Debye model of harmonic solids
HDA reveals the highest Debye temperature among the studied ice phases,
which is in full agreement with the lowest Debye level in the generalised
density of states derived from time--of--flight neutron scattering
experiments.
The elastic results at low $T$ indicate the presence of an excess
of modes in HDA, which do not obey the Bose statistics.
\end{abstract}


\maketitle
{\bf I. INTRODUCTION}\\

Water, although comprising one of the most simple molecules,
reveals extremely complicated properties in its condensed
states.
In the solid state for example, there exist not only
twelve well known crystalline phases but also a number
of disordered structures \cite{Hobbs-Book,Petrenko-Book}.
Two of the disordered structures are known as high--density amorphous (HDA)
and low--density amorphous (LDA) ice, characterised by a molecular
density of $\rho\approx39$~molec./nm$^3$ and $\rho\approx31$~molec./nm$^3$,
respectively \cite{Mishima-Nature-1984,Mishima-Nature-1985}.
The origin of a number of different disordered
phases within a single substance, often referred to as amorphous
polymorphism, has not been understood, yet.
It has been conjectured on the basis of computer simulation results
that HDA and LDA might represent the supercooled glassy structures
of two different liquid phases of water
\cite{Poole-PRE-1993,Poole-PRL-1994,Debenedetti-Book,Mishima-Nature-1998,Stanley-PCCP-2000}.
Obviously, this scenario implies the existence of a
glass--transition designated by a glass--transition temperature
$T_{\rm g}$.

From the experimental point of view, there is a number of
results backing the findings from computer simulations.
Structures equivalent to HDA and LDA can be obtained
from the liquid by extremely fast quenching
of $\mu$m--droplets or from the gas phase by vapour deposition
\cite{Narten-JCP-1976,Mayer-Nature-1982,Mayer-JCP-1983,Mishima-JCP-2001,Comment-0}.
As far as LDA--type structures are concerned, they all transform
exothermally to a crystalline cubic phase (ice I$_{\rm c}$)
at $T\approx 150-160$~K on an endothermal plateau setting in
at $T\approx 135$~K, which has been observed by differential scan
calorimetry (DSC) and other calorimatric experiments
\cite{Handa-JCP-1986,Johari-Nature-1987,Handa-JPC-1988,Hallbrucker-JPC-1989,Hallbrucker-PM-1989,Johari-JCP-1990,Johari-JPC-1990,Mayer-JMS-1991,Johari-JCP-1991,Salzmann-PCCP-2003,Mishima-JCP-2004}.
The very onset of the endothermal contribution has been interpreted
as water's $T_{\rm g}$ in literature, whereby in LDA
the lowest $T_{\rm g}\approx 124$~K of all low--density amorphous
structures is discussed \cite{Handa-JPC-1988}.

A comparable $T_{\rm g}$ has been deduced from DSC studies
of different aqueous solutions, whose specific $T_{\rm g}$s
extrapolate upon dilution toward the single value of 135~K
\cite{Angell-AC-1976,Angell-CR-2002}.
Finally, from observation of D$_2$O--H$_2$O interlayer mixing
upon heating of LDA--type vapour deposited
water a higher molecular mobility at $T\approx 150$~K
has been concluded \cite{Smith-Nature-1999,Smith-CP-2000}.

However, these experimental results do not establish unequivocally
a glass--transition or a glass--transition temperature in water.
For each of the cited experiments there can be found at least one experiment
{\it reported in literature}
whose results either contrast the glass transition scenario
or give an alternative explanation for the observed phenomena.
For example, endothermal transitions have been observed at the same thermodynamic
conditions in transformations of high--density crystalline
ice phases to ice I$_{\rm c}$.
They are understood as entropy driven order~$\rightarrow$~disorder
transitions in the proton sublattice of the phases
\cite{Handa-JDP-1987,Handa-CJC-1988,Salzmann-PCCP-2003}
a feature equally detected in ice~I$_{\rm c}$
\cite{Yamamuro-JPCS-1987}.
In this context, it is noteworthy that recently well established
DSC results on apparent LDA samples
\cite{Hallbrucker-JPC-1989,Johari-JCP-1990,Mayer-JMS-1991,Johari-JCP-1991}
had to be reinterpreted as such a reversible
order~$\rightarrow$~disorder transition in the proton sublattice
of ice~XII \cite{Salzmann-PCCP-2003}.
The very low $T_{\rm g}\approx 124$~K reported in \cite{Handa-JPC-1988}
remains to be reexamined \cite{Salzmann-PCCP-2003}.
Consequently, one may ask what is the unique dynamic feature by which,
on the one hand, the endothermic plateau in the amorphous systems
is interpreted as a glass--transition but, on the other hand,
it is identifiied in the crystalline structures as a disorder
phenomenon of the proton sublattice only.
Whether the recently observed endothermic effect in pressure
dependent experiments on emulsified amorphous samples
is a unique feature of the disordered system
{\it requires equally a closer examination} \cite{Mishima-JCP-2004}.

In contrast to the D$_2$O--H$_2$O interlayer mixing effect
\cite{Smith-Nature-1999,Smith-CP-2000}
ion diffusion experiments have not shown any higher mobility
of guest molecules in the amorphous ice matrix
\cite{Tsekouras-PRL-1998}
neither have isotopic exchange measurements shown any
translational motion of water molecules
\cite{Fisher-JPC-1995}.
Instead, they have given the picture of a defect promoted proton
mobility responsible for the endothermic transition,
a feature directly detectable by Nuclear Magnetic Resonance
in crystalline ice and clathrate hydrates
\cite{Fujara-JCP-1988,Kirschgen-PCCP-2003}.

Recently, models of a so called shadow
glass--transition have been proposed that explain the thermal
anomalies observed at $T\approx135$~K with a real but
experimentally unaccessible glass--transition at
$T_{\rm g}\approx 165$~K \cite{Velikov-Science-2001}.
Whether this shadow glass--transition could be indeed applicable to water
is the issue of current debates \cite{Johari-JCP-2003,Yue-Nature-2004}.

Another idiosyncratic feature which has not been satisfactorily explained,
yet, is the inelastic response of the amorphous ice structures.
Dynamic properties of bulk HDA, LDA and LDA--type structures
differ strongly from established features of supercooled glassy systems
\cite{Sokolov-PRL-1997,Sette-Science-1998}.
The dynamic response of LDA does not exhibit any phonon damping
\cite{Schober-PRL-2000,Koza-PhD-2001},
any remarkable Boson--peak
\cite{Koza-PhD-2001,Koza-ILL-Report,Schober-PB-1998,Yamamuro-JCP-2001},
and any two--level systems (TLS)
\cite{Agladze-PRL-1998,Agladze-EPL-2001,Agladze-PB-2002}.
Consequently, its spectral density and thermal conductivity
are reminiscent of a harmonic crystalline state
\cite{Andersson-PRB-2002}.
In the case of HDA, this departure from the characteristics
of glassy states is apparently less pronounced, since TLS modes
have been unequivocally observed in optical absorption
experiments \cite{Agladze-PRL-1998,Agladze-EPL-2001,Agladze-PB-2002}.
Similar features are observed by inelastic neutron experiments
\cite{Svensson-PB-1994,Tulk-APA-2002,Tse-PRL-2000},
where an excess of modes in the density of vibrational
states is reported.
However, the detailed phonon dispersion and also the spectral
density at $T>40$~K of HDA are identified as crystal--like
in other experiments pointing at an intriguing high degree
of short--range order \cite{Koza-PRB-2004,Schober-PRL-2000}.

In the light of the above outlined experimental facts
the idea of HDA and LDA being strongly disordered crystalline systems not
thermodynamically connected to water's liquid phase \cite{Johari-PCCP-2000},
may at first glance seem rather tempting.
Such a nano--crystalline scenario has been discussed for water
\cite{Johari-PCCP-2000} but also for
other systems showing amorphous polymorphism
\cite{Tarjus-JPCM-2003}.
A discontinuity between the liquid state, vapour deposited and
hyper--quenched amorphous water on the one hand and LDA
on the other hand has been indeed postulated based on theoretical
concepts and results from spectroscopic
experiments and computer simulations
\cite{Tse-Nature-1999,Klug-PRL-1999,Johari-JCP-2000,Shpakov-PRL-2002}.
Giving the subject some further thinking it becomes, however,
obvious that a nano-crystalline scenario does not provide a more
stringent explanation for the absence of glassy features
in the dynamics. The inherently high disorder within the nano-crystals
should leave a clear trace in the low-frequency dynamics not
too different from that of a glassy state.

To advance the question further it is obviously indispensable to
characterize the fundamental, microscopic relaxation processes
responsible for the endothermic plateau in DSC experiments.
This characterization must be carried out both in time and space.
While order--disorder phenomena and the glass relaxation may take place
on similar time scales the spatial correlations are different.
In particular, the spatial patterns should allow to discern the hypothetical
relaxation channels of the glass--transition from processes observed
during the transitions of high--density crystalline
structures to ice~I$_{\rm c}$.

The present work represents our first attempt of scrutinizing
the glass--transition hypothesis of water with a supposed
$T_{\rm g}\approx 135$~K on a microscopic scale
by neutron backscattering and time--of--flight techniques, i.e. by
spectroscopic techniques sampling time and spatial correlations.
Both techniques have been extensively used for studying glass
transition phenomena and proved to be substantial for their
understanding.
Both, time--of--flight and backscattering spectroscopy offer
the opportunity of studying the microscopic dynamics of a sample
directly as an energy resolved response.
In addition, neutron backscattering can be also used to exploit
the elastic scattering within a narrow energy range $\sim 1$~$\mu$eV
from which important dynamic properties can be deduced
\cite{Springer-Book,Bee-Book}.
This enables us, in principle, to characterise relaxation processes
on a molecular scale during the apparent glass--transition.

All experiments performed reveal that the properties of
HDA and LDA follow a temperature dependence
in complete accordance with the harmonic theory of
the solid state \cite{Bee-Book,Lovesey-Book}.
Beyond the low--temperature limit, where zero--point
oscillations are predominant, the molecular mean--square
displacements of HDA and LDA are proportional
to the temperature increase.
This behaviour is expected if the degrees of freedom,
and thus the spectral density of the sample in its different
states is determined by harmonic modes only.
The very good conformity with the mean--square
displacements deduced from the phonon densities of states
of HDA and LDA reinforces this scenario.
In the temperature regime of the transitions
HDA~$\rightarrow$~LDA and LDA~$\rightarrow$~ice I$_{\rm c}$
comprising the apparent glass--transition with
$T_{\rm g}\approx 135$~K
no higher mobility of water molecules, e.g., in terms of
translational diffusion, is observed on a time scale
shorter than some nano--seconds.
From experiments on coherently scattering samples
the temperature $T\approx 135$~K is identified as
the onset of a recrystallisation of LDA into ice I$_{\rm c}$
\cite{Geil-PCCP-2004}.
In addition, an excess of modes in HDA, that however does
not follow the Bose--statistics valid for harmonic vibrations,
is clearly indicated in the values of the zero--point oscillation.

To bring these findings clearly forward the present paper
is structured in the following way.
The next two chapters render some useful details on the applied
neutron scattering techniques, the samples and experimental procedure,
and introduce some observables (Debye--Waller factor,
mean--square displacement, density of states, velocity
of sound), to which we refer throughout the text.
In section IV, we present and discuss the experimental data in view of
the studied transitions (HDA~$\rightarrow$~LDA,
LDA~$\rightarrow$~I$_{\rm c}$, and the glass--transition),
the Debye--Waller factor and mean--square displacement,
and the excess of modes in HDA.
All results obtained are summarised in section V.\\

{\bf II. EXPERIMENTAL}\\

To meet the requirement of high energy resolution the experiments
have been performed on the neutron backscattering
spectrometers IN13 and IN16 and the time--of--flight
spectrometer IN6 at the Institut Laue Langevin in Grenoble,
France \cite{yellow-book}.
The principles of the backscattering spectrometers are based on
the neutron beam monochromatization and the energy analysis of
the scattered beam by single crystal Bragg reflection
\begin{equation}
n\cdot \lambda = 2d\sin\left(\Theta\right)
\label{eq_one}
\end{equation}
in backscattering geometry $\Theta\approx 90^{\circ}$
\cite{Springer-Book,Bee-Book}.
Using the CaF$_2$(422) reflection on IN13 and the Si(111) on IN16
the incident neutron energies ($E$) and energy resolutions
($\Delta E$) of $E=16.5$~meV, $\Delta E=8\mu$eV and
$E=2$~meV, $\Delta E= 1 \mu$eV are obtained, respectively.
Beyond the differences in incident energy and energy resolution
these spectrometers are optimised for sampling complementary
$Q$--ranges.
IN16 with $Q<2.1$\AA $^{-1}$ is best utilised for e.g. long
range diffusion processes with a correlation length of
$r>3$~\AA .
IN13 with $Q<5.5$\AA $^{-1}$ is optimised for short range
diffusion with $r>1$~\AA .
Thus, with both instruments we are able to cover a spatial
range stretching from correlations within the proton sublattice
of water up to its intermolecular distances.

In general, backscattering spectrometers can be utilised
in two different modes.
Firstly, in the elastic scan mode, the incident neutron energy
is kept constant to permanently meet the Bragg condition
of the analyser given by eq.~\ref{eq_one}.
The elastic scan mode samples changes of the elastic intensity
within $\Delta E$ due to dynamic and relaxation processes
which may be induced by changing experimental parameters
like temperature in the present case.
Secondly, in the energy scan mode, the incident neutron energy
is varied in a well defined, systematic way within a narrow energy range
$\delta E$, e.g., $\delta E\approx \pm 10\mu$eV on IN16.
In the energy scan mode, dynamic and relaxation processes can be,
in principle, characterised quantitatively.
Both, energy and elastic scan mode, were applied on the spectrometer
IN16 whereas IN13 was used in the elastic scan mode exclusively.

The time--of--flight spectrometer IN6 was used with incident
energies of 3.1~meV ($\Delta E= 80$~$\mu$eV) and 4.8~meV
($\Delta E= 150$~$\mu$eV).
The sampled $Q$--ranges of 0.3--2.1~\AA $^{-1}$ (3.1~meV) and
0.3--2.6~\AA $^{-1}$ (4.8~meV) matched the $Q$ regime of the IN16
measurements.
However, the energies probed by the instrument IN6 correspond to
dynamic processes on a time scale of $\tau < 50$~pico--seconds.

\begin{figure}
\begin{center}
\includegraphics[angle=0,width=140mm]{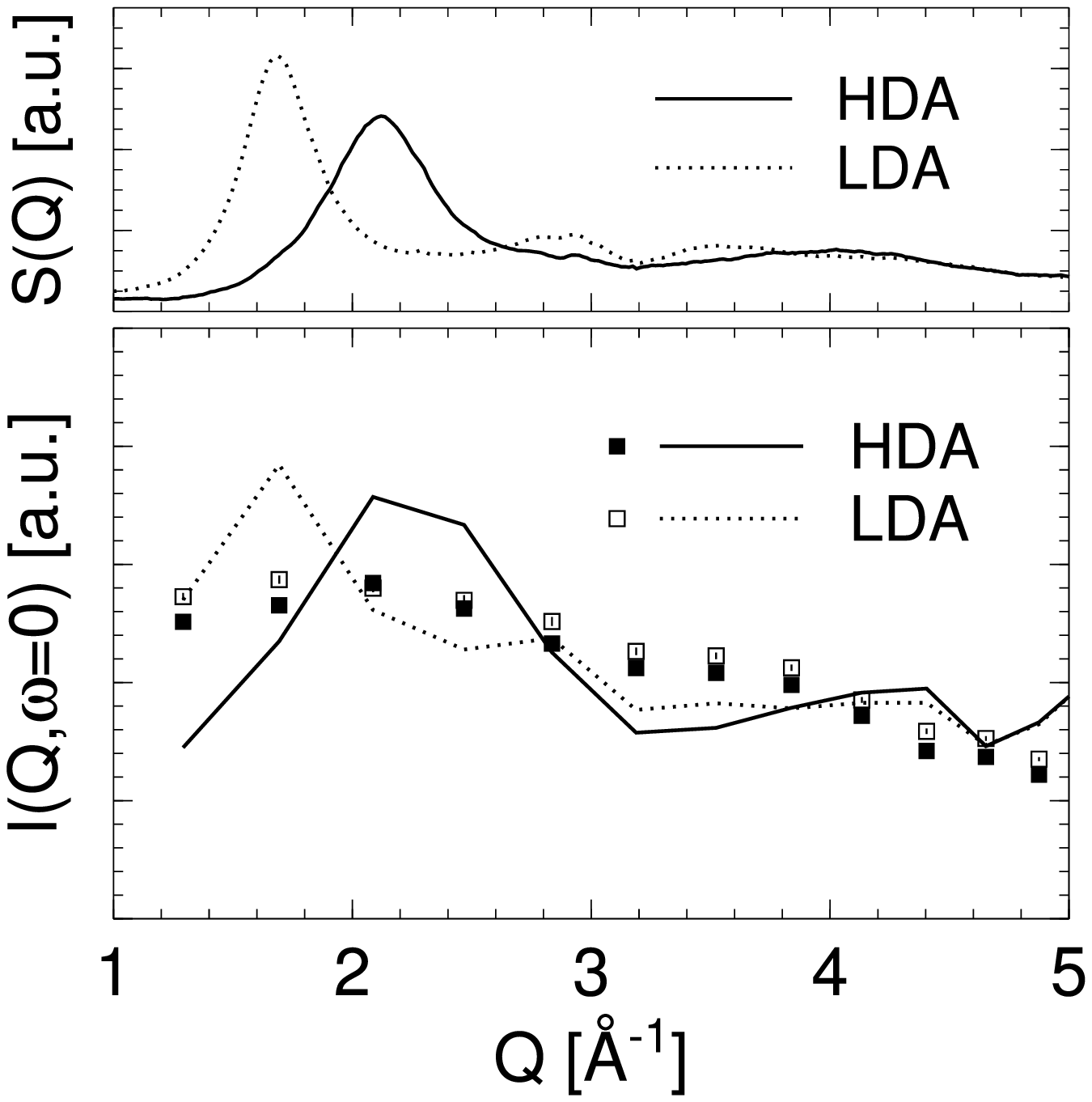}
\end{center}
\caption{Bottom: Elastic intensity $I(Q,\omega=0)$ of the amorphous samples
measured on the spectrometer IN13 at $T\approx 2$~K.
Solid and dotted lines represent the HDA and LDA structures studied with
the deuterated sample S2, respectively.
$\blacksquare$ and $\square$ represent the HDA and LDA structures studied with
the partially protonated sample S1, respectively.
Top: For comparison static structure factor of the deuterated
HDA and LDA samples measured with high resolution on the
diffractometer D20.}
\label{fig_01}
\end{figure}

In detail, we studied two fully protonated samples on IN16,
one sample, S3, in elastic scan and a second, S4, in energy scan
mode, and two samples on IN13, sample S2 fully deuterated
and sample S1 partially protonated with $40\%$~vol. H$_2$O.
Three fully deuterated samples were measured on IN6.
The deuteration was employed for a better control
of the sample state via the pronounced coherent scattering
contribution of D$_2$O to the signal and for an enhanced
signal contribution from the Oxygene atoms.
Figure~\ref{fig_01}  shows the elastic intensity
$I(Q,\omega=0)$ of the samples S1 and S2
in the HDA and the LDA state in comparison to the structure factor
$S(Q)=\int S(Q,\omega)\rm{d}\omega$ of fully deuterated HDA and LDA
measured with high resolution on the diffractometer D20
at the Institut Laue Langevin \cite{Koza-JPCM-2003}.
The coherent signal of the partially protonated sample
is, as expected, strongly suppressed and the scattering
characteristics are reminiscent of the incoherent contribution.

All samples were prepared by compression of crystalline ice I$_{\rm h}$
at $T\approx 77$~K (in liquid N$_2$) in a piston--cylinder apparatus
up to $p\approx 18$~kbar \cite{Koza-PRL-2000}.
The formed HDA was recovered at ambient pressure, filled
into Aluminum holders used as standard in neutron scattering
and placed into precooled ($T=75$~K) standard cryostats.
After remaining N$_2$ had been carefully removed at $T=78$~K
the samples were cooled down to $T=2$~K.
Table \ref{tab_one} gives details on the thermal treatment and
thermal cycling applied to the samples during the experiments.
Data were accumulated on IN13 for 60~min per point with the
samples gradually heated corresponding with a rate
of 3.5~K/60~min.
IN16 elastic scan data were accumulated for 2~min
with a heating rate of 1~K/2~min.
IN16 inelastic scan data were accumulated for $2\times90$~min
with a heating step of 2~K in the range of the supposed
glass transition, whereby {\it in situ} data updates were performed
every 10~min giving, within the data statistics, identical results
as obtained after 90~min acquisition time.
As for the LDA state the thermal treatments resulted in phases
which had been annealed for 2~h at 125~K plus 1~h at 128.5~K on IN13
and for 1~h at 130~K and a preceeding slow heating on IN16
before the regime of the apparent glass transition was reached.
Thus, annealing times and heating rates corresponded with
treatments performed in DSC experiments on water's $T_{\rm g}$
reported in literature.

Samples measured at IN6 were subject to a comparable thermal cycling
procedure whereby the LDA structure was annealed at 127~K
and ice~I$_{\rm c}$ was annealed at 160~K.
Measurements following the glass--transition have been performed
at the temperatures $T=127$~K for $4\times 60$~min.,
at $T=132$~K for $4\times 60$~min., at $T=137$~K for $5\times 30$~min.,
at $T=140$~K for $6\times 30$~min., at $T=143$~K for $4\times 30$~min.,
at $T=147$~K for $30$~min. and at $T=155$~K for $30$~min.
before annealing the formed ice~I$_{\rm c}$ at 160~K.
Long time measurements of 4--6~hours were carried out with HDA
at 2~K, 20~K, 40~K, 60~K and 80~K, with LDA at 2~K, 20~K, 40~K, 60~K, 80~K
and 127~K, and ice~I$_{\rm c}$ at 127~K and 160~K.

\begin{table}
\caption{\label{tab_one}
Thermal cycling applied to the samples S1 and S2 (IN13) and S3 and S4 (IN16).
The phases given below indicate the states in which the samples existed
when heat treatment was started at 2~K for the measurements.
Please note that S1 was measured twice within the stability
range of HDA.}
\begin{ruledtabular}
\begin{tabular}{ccccc}
phase & sample S1 & sample S2 & sample S3 & sample S4 \\
\hline
HDA             & 2$\rightarrow$65$\rightarrow$2$\rightarrow$125~K &
2$\rightarrow$125~K & 2$\rightarrow$130~K & 2$\rightarrow$160~K \\
LDA             & 2$\rightarrow$170~K & 2$\rightarrow$170~K
& 2$\rightarrow$260~K & -- \\
ice I$_{\rm c}$ & 2$\rightarrow$170~K & -- & -- & -- \\
\end{tabular}
\end{ruledtabular}
\end{table}

Standard data corrections were performed, accounting for empty
container and background scattering, self-shielding, absorption
effects of the samples and for efficiencies of the instrument detectors.
However, due to the scattering power of the samples (optimised
for multiple scattering behaviour) and the negligible absorption of
the scatterers only minor perturbations of the signal were observed.
Nevertheless, the elastic signal measured in the low--$Q$ range
on the spectrometer IN16 encounters a lower sensitivity to temperature
changes in comparison to the results taken on IN13.
This feature is well reported in literature and associated with
contribution from multiple scattering processes
\cite{Wuttke-PRE-2000}.
Depending on the information which is to be extracted from the
experimental data, different normalisation standards have to be used.
These standards are explicitely described in the text below.

For a clear presentation of the extensive amount of data
some data sets have been regrouped to give statistical errors
of the order of the data symbols.
Error bars are plotted with the data in the figures.
The grouping is explicitly indicated in the figure captions
for data whose interpretation exceeds a pure qualitativ description.
All fitting routines were performed with ungrouped data sets.
The simplex minimization algorithm was applied having
taken into account the x--grid and y--error of the fitted data.

Please note, that we focus primarily on the backscattering measurements
due to the superior resolution of the instruments, when discussing
potential relaxation processes during the phase transitions HDA
to LDA and LDA to ice~I$_{\rm c}$.\\

{\bf III. OBSERVABLES IN NEUTRON SCATTERING }\\

For the convenience of the reader let us summarise some basic principles
of neutron scattering as they are needed for the comprehension
of the data analysis and the discussion.
A more detailed introduction into this field can be found
in textbooks on neutron scattering
\cite{Springer-Book,Bee-Book,Lovesey-Book}.
At the beginning we restrict our consideration to a harmonic system
whose scattering characteristic is assumed to be isotropic.
These terms are well met by amorphous solids.
For such a system the intensity measured
in an elastic scan mode neutron scattering experiment can be written as
\begin{eqnarray}
S(Q,\omega=0) & = &  \exp{(W(Q,T))}\cdot S(Q)
\label{eq_two}\\
W(Q,T)          & = &  -Q^2\left<u^2(T)\right>
\label{eq_three}
\end{eqnarray}
with $W(Q,T)$ the Debye--Waller factor, $\left<u^2(T)\right>$
the mean--square displacement of the scatterers,
and $S(Q)$ the static structure factor, which may be approximated
for incoherent scattering by a scaling factor $(2\pi N)^{-1}$ representing the
number of scatterers in the sample
\cite{Comment-0b}.
Harmonicity implies that $\left<u^2(T)\right>$ can be related to
the density of states denoted as $G(\omega)$ as
\begin{equation}
\left<u^2(T)\right> = \frac{1}{6} \frac{\hbar^2}{M}\int
\frac{G(\omega)}{\hbar\omega}\coth{\left(\frac{\hbar\omega}{2k_{\rm B}T}
\right)}{\rm d}\omega
\hspace{2ex},
\label{eq_four}
\end{equation}
$M$ representing the mass of the scatterers \cite{Comment-0c}.
Taking the low and high temperature limits into consideration
$\left<u^2(T)\right>$ can be asymptotically approximated by
the moments of energy
$\left<\omega^{-1}\right>$ und $\left<\omega^{-2}\right>$ as
\begin{eqnarray}
\left<u^2(T\rightarrow 0)\right> & = &
\frac{\hbar^2}{6M} \cdot \int\frac{G(\omega)}
{\hbar\omega}{\rm d}\omega
\label{eq_five}
\hspace*{2ex}, \\
\left<u^2(T\rightarrow \infty)\right> & = &
\frac{\hbar^2}{3M} \cdot k_{\rm B}T \cdot \int
\frac{G(\omega)}{\hbar^2\omega^2}{\rm d}\omega
\label{eq_six}
\hspace*{2ex}.
\end{eqnarray}
As a consequence, the mean--square displacement is expected to cross over
from a constant value at low temperatures to a $\propto T$ behaviour upon heating
leading to the relation
\begin{equation}
W(Q,T)\propto -Q^2\cdot T
\label{eq_seven}
\hspace*{2ex}.
\end{equation}
Please note, that $\left<u^2(T\rightarrow 0)\right>$ measured
in neutron scattering experiments is not the absolute zero--point
oscillation of the scatterers.
Its value depends on the incident neutron energy.
This energy sets a limit to the energy range on which the
spectral density of the scatterers is sampled at $T\rightarrow0$~K.
However, $\left<u^2(T\rightarrow 0)\right>$ denotes a well defined
quantity which makes possible to draw comparative conclusions upon
the properties of different sample states under identical experimental
conditions.

Going back to eq.~\ref{eq_two}, we can quantitatively determine
two physical effects which can induce changes in the observable elastic
intensity $S(Q,\omega=0)$.
Firstly, the structure of the sample detectable as $S(Q)$ can change
due to phase transformations.
Secondly, the spectral density $G(\omega)$ can alter and influence
both the low and high temperature properties of $S(Q,\omega)$  via
eqs.~\ref{eq_five} and \ref{eq_six}, respectively.
From the experimental point of view neutron scattering offers
the opportunity of discerning these two points by utilising,
on the one hand, incoherent scatterers (protonated samples S1, S3 and S4)
to study changes in $G(\omega)$ and, on the other hand,
coherent scatterers (deuterated samples S2) to measure also changes
in $S(Q)$.

Finally, utilising the Debye approximation for $G(\omega)$
\cite{Ashkroft-Book}
\begin{equation}
G(\omega)   =   \frac{1}{2\pi^2}\cdot\frac{V}{N}\cdot\frac{3}{\bar{v}^3}
\cdot\omega^2 \\
\label{eq_eight}
\end{equation}
with $\bar{v}$ the average velocity of sound
\begin{equation}
\frac{3}{\bar{v}^3} = \frac{1}{v^{3}_{l}} + \frac{2}{v^{3}_{t}}
\label{eq_nine}
\end{equation}
the observables $G(\omega)$ and $\left<u^2(T)\right>$ given in
eqs.~\ref{eq_five} and \ref{eq_six}
can be compared qualitatively and quantitatively with results
from other experimental techniques accessing the velocity of sound.
Within the Debye model a characteristic quantity is the
Debye temperature $T_{\rm D}$
\begin{equation}
T_{\rm D}^3 = 2\pi^2\cdot\left(\frac{\hbar}{k_{\rm B}}\right)^3\cdot
\frac{3sN}{V}\cdot\bar{v}^3
\hspace*{2ex},
\label{eq_ten}
\end{equation}
with $3sN/V$ the density of vibrational modes of the system,
which can be also extracted by numerical techniques
from eq.~\ref{eq_four} as
\begin{equation}
\left<u^2(T)\right> = \frac{3}{2} \frac{\hbar^2}{Mk_{\rm B}}
\cdot\frac{T^2}{T_{\rm D}^3}\cdot\int_0^{T_{\rm D}/T}
x\cdot\coth{(x/2)}\cdot{\rm d}x
\hspace*{2ex}.
\label{eq_eleven}
\end{equation}
\\

{\bf IV. RESULTS AND DISCUSSION}\\

{\bf IV.a. Phase transformations and glass--transition}\\

\begin{figure}
\begin{center}
\includegraphics[angle=0,width=140mm]{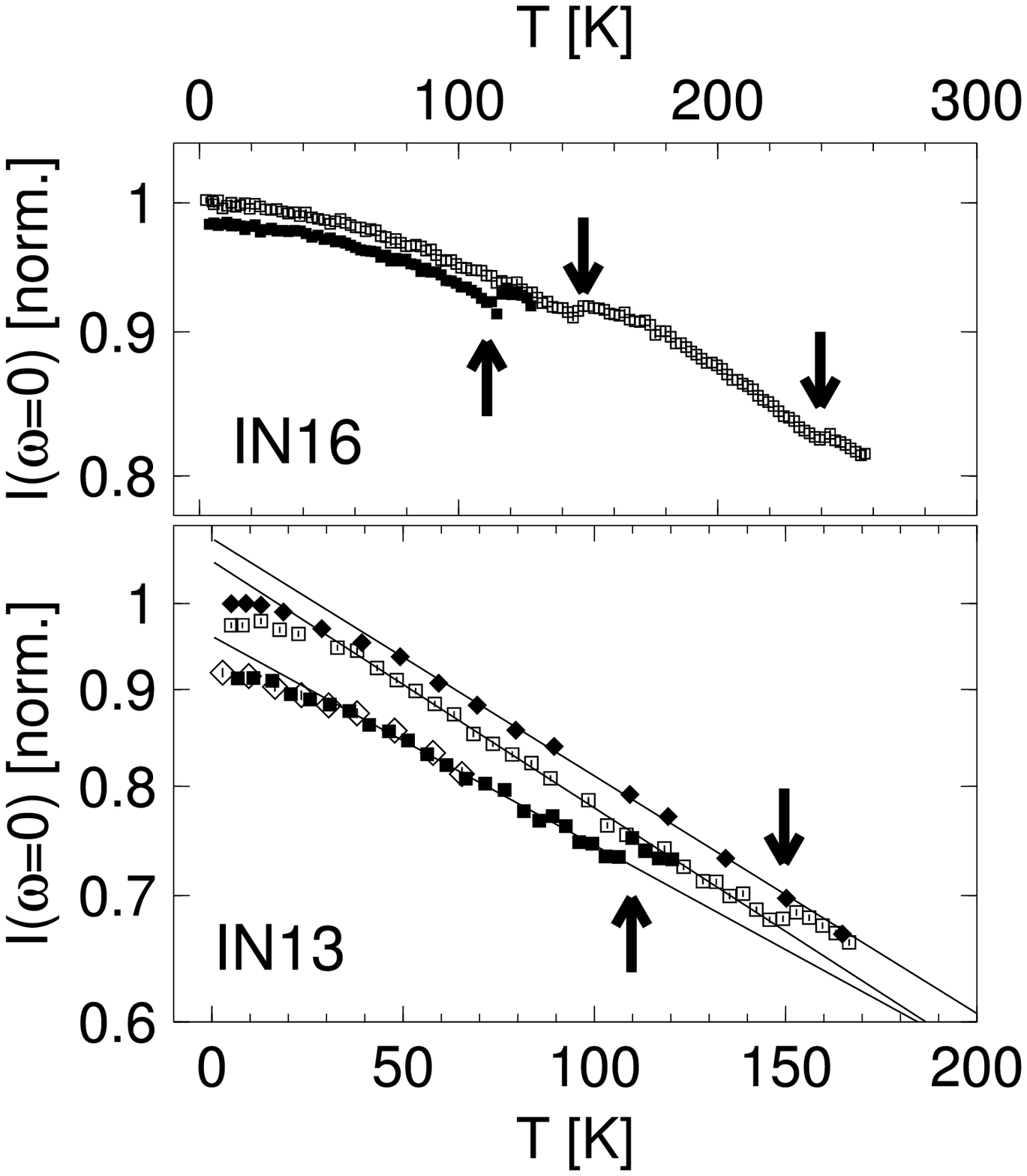}
\end{center}
\caption{Temperature dependence of the elastic intensity
$I(\omega=0)$ measured with the partially protonated sample S1 on IN13
and the fully protonated sample S3 on IN16.
The signal is integrated over the accessible $Q$--range of the instruments.
Data on HDA are represented as $\lozenge$ (IN13) and $\blacksquare$
(IN13 and IN16), on LDA as $\square$ (IN13 and IN16),
and on cubic ice I$_{\rm c}$ as $\blacklozenge$.
Details on the thermal treatment of the samples are given in the text.
The intensity is normalised to the $I(\omega=0)$ at $T\approx 2$~K
of I$_{\rm c}$ on IN13 and of LDA on IN16.
Upward arrows indicate the transformation HDA~$\rightarrow$~LDA
at $T\approx110$~K, downward arrows indicate the transformation
LDA~$\rightarrow$~I$_{\rm c}$ at $T\approx 150$~K and from cubic
to hexagonal ice at $T\approx240$~K measured on IN16 exclusively.
Solid lines stress the linear behaviour of the intensity expected
for a harmonic solid at sufficiently high temperature.}
\label{fig_02}
\end{figure}

To give an overview on the temperature dependence of the elastic
signal and its changes during the thermal treatment,
Fig.~\ref{fig_02} reports on the intensity $I(\omega=0)$
measured in elastic scans with S1 on IN13 and S3 on IN16.
$I(\omega=0)$ is the intensity integrated over the accessible
$Q$ range of the instruments.
The data sets are normalised to the intensity detected at $T=2$~K
in the crystalline cubic phase I$_{\rm c}$ (S1, IN13) and LDA (S3, IN16).
The phase transformations HDA~$\rightarrow$~LDA and
LDA~$\rightarrow$~I$_{\rm c}$
can be identified in both samples as rather sharp intensity gains
in the elastic response at $T\approx110$~K and $T\approx150$~K.
They are indicated by vertical arrows.
Additionally, a small step in intensity at $T\approx240$~K
indicates the formation of the hexagonal crystal
from ice I$_{\rm c}$ as reported in reference \cite{Handa-JCP-1986,comment-iceIh}.

At sufficiently high $T$ ($\ge 40$~K),
the intensity of all phases decreases linearly upon heating
in the logarithmic presentation of Fig.~\ref{fig_02}.
This behaviour is stressed by solid lines.
Neither in the $T$ range of HDA$\rightarrow$LDA nor
in the $T$ regime of the LDA$\rightarrow$I$_{\rm c}$
and, thus, the assumed glass--transition,
a pronounced drop off in $I(\omega=0)$ can be detected.
Such a drop off would be a mandatory fingerprint for
an enhancement in molecular mobility of the sample
(eq.~\ref{eq_two}) on time scales less than a few
nano--seconds.

A missing enhancement in molecular mobility
is also confirmed by energy scan measurements
with the incoherently scattering sample S4 on IN16.
Spectra measured at different temperatures are reported
in Fig.~\ref{fig_08}.
To compensate for the here nonrelevant temperature dependence
of the elastic intensity the data are normalised to the maximum
intensity at the respective temperatures.
The identity of all spectra demonstrates that within the instrumental
resolution of 1~$\mu$eV no relaxation processes are present
in the sample.
From the comparison of the different data to the results measured
at 2~K it is evident that the spectra do not change markedly their shape
in the energy window of $\pm$10~$\mu$eV,
regardless of the state of the sample (The contribution of phonons
in this energy window is too weak to leave a trace in the temperature
dependence.)
In accordance with the IN16 data, no change of the inelastic
response apart from the normal harmonic phonon contribution can be
observed in the IN6 experiments at temperatures
prior to the recrystallisation of the amorphous samples.
We may conclude that according to eqs.~\ref{eq_two} and \ref{eq_three}
there is no evidence for a development of additional decay channels
for the elastic intensity at the assumed glass transition
with $T_{\rm g}\approx 135$~K.
There is no evidence of relaxation behaviour
in the course of the transformations HDA~$\rightarrow$~LDA and
LDA~$\rightarrow$~I$_{\rm c}$, either.
Consequently, the presence of relaxation processes
like translational diffusion over intermolecular distances
on a time scale shorter than some nano--seconds can be excluded.

\begin{figure}
\begin{center}
\includegraphics[angle=0,width=140mm]{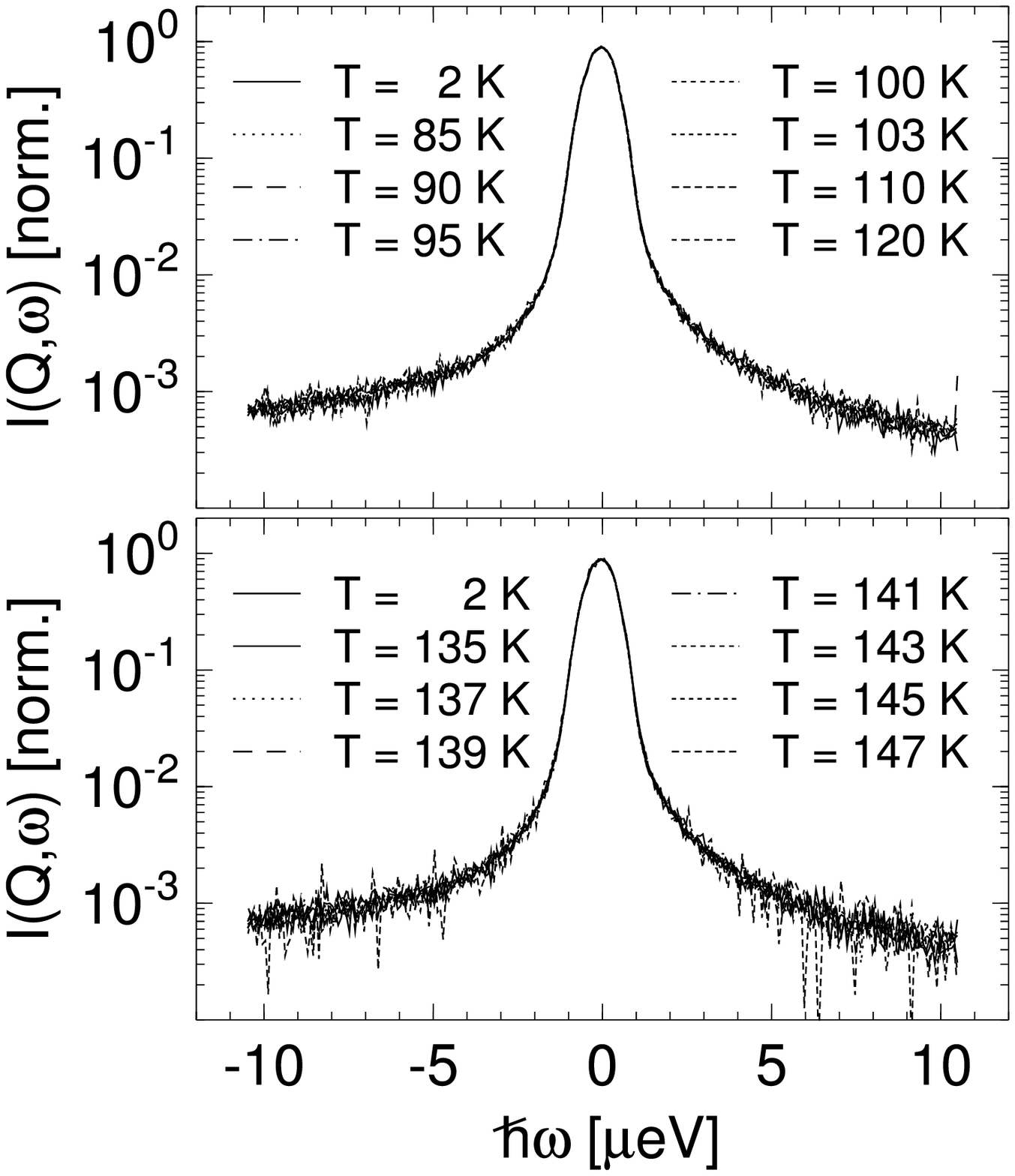}
\end{center}
\caption{Energy scans $I(Q,\omega)$ of the fully protonated
sample S4 measured on IN16.
All spectra are normalised to the maximum intensity determined
at the respective temperatures that are indicated in the figures.
Top figure reports spectra measured in the course of the
HDA~$\rightarrow$~LDA transformation, bottom figure reports
spectra measured in the temperature range of the apparent glass
transition.
Also plotted are data obtained at $T\approx2$~K representing
the instrumental resolution.
Please note, that there is no difference between the $I(Q,\omega)$
at different temperatures. }
\label{fig_08}
\end{figure}

\begin{figure}
\begin{center}
\includegraphics[angle=0,width=140mm]{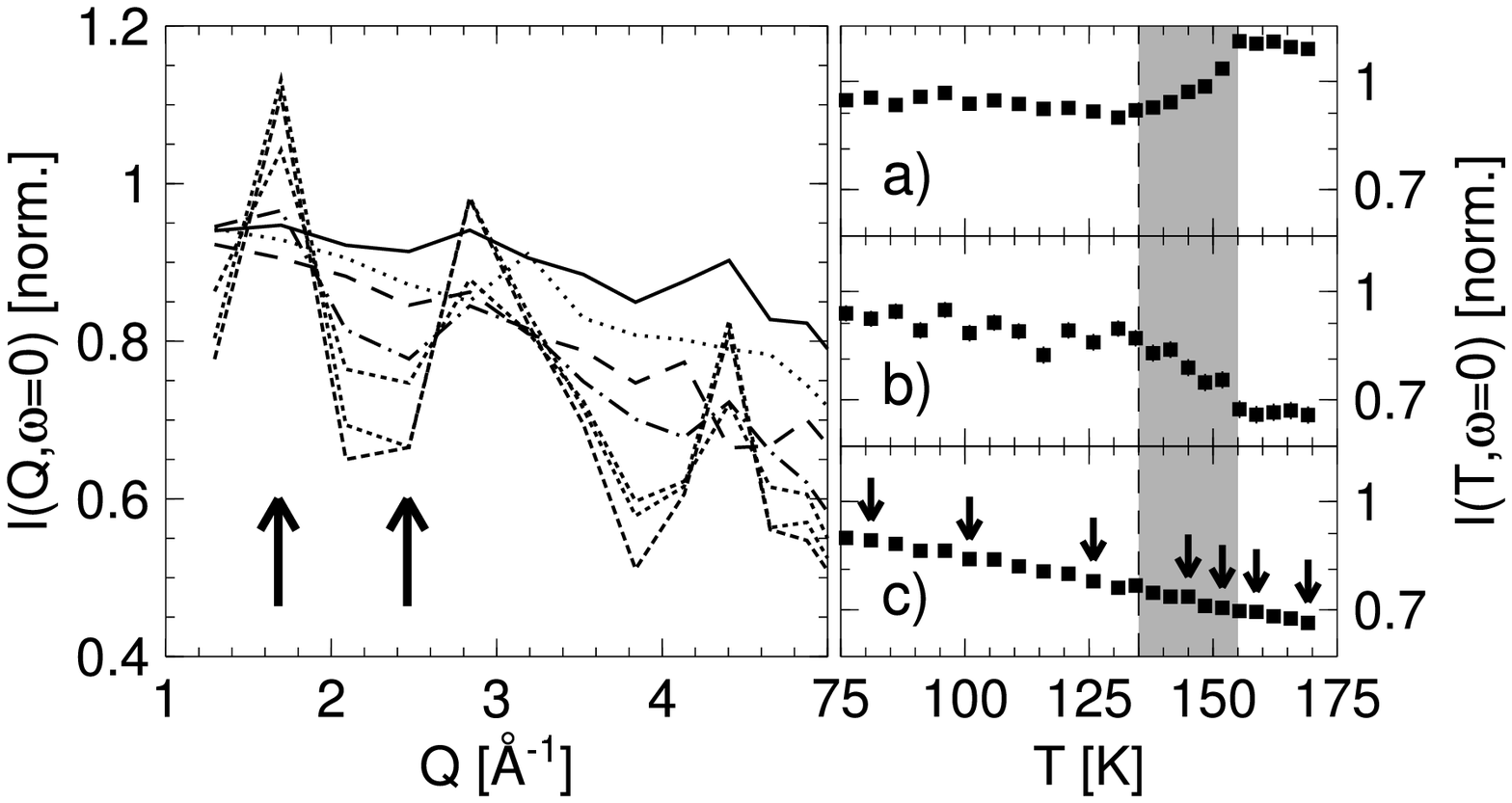}
\end{center}
\caption{Left: Temperature evolution of the elastic intensity
$I(Q,\omega=0)$ during the transformation
of LDA into ice~I$_{\rm c}$
measured on the spectrometer IN13 between
$T\approx 80$~K and $T\approx 170$~K.
The nominal temperatures are indicated by arrows in
figure~c in the right panel.
The two arrows indicate the $Q$--values at which the temperature
dependence $I(T,\omega=0)$ is inspected in detail in
figures a and b in the right panel.
Right: Elastic intensity  $I(T,\omega=0)$ measured at
$Q=1.7$~\AA $^{-1}$ in figure a, at $Q=2.4$~\AA $^{-1}$ in figure b,
and integrated over the accessible $Q$--range in figure~c.
Dashed vertical line stresses the supposed glass transition
and the grey shaded area indicates the transition regime
of LDA to ice~I$_{\rm c}$}
\label{fig_04}
\end{figure}

Before we will discuss the Debye--Waller factor in detail
and, thus, the potential relaxation processes on shorter
length scales, it is important to establish the regime
of stability of the amorphous structures.
We focus here primarily on the supposed glass transition and the
LDA~$\rightarrow$~I$_{\rm c}$ transformation.
Fig.~\ref{fig_04} reports on the elastic intensity
$I(Q,\omega=0)$ measured with S2 on IN13 at different temperatures
which correspond to the regime of the glass transition.
The temperatures are indicated by vertical arrows
in Figure~\ref{fig_04}~c.
Here, $I(Q,\omega=0)$ is normalised to $I(Q,\omega=0)$ of LDA
determined at 2~K.
It gives, therefore, the relative changes before and after
the conversion of LDA into I$_{\rm c}$.
The two arrows indicate the $Q$ values at which the temperature
dependent intensity $I(T,\omega=0)$ is plotted in
Figure~\ref{fig_04}~a and b.
As it is expected from a harmonic system $I(Q,\omega=0)$
exhibits up to 130~K only a loss of intensity due
to the Debye--Waller factor (eq.~\ref{eq_three}).
Whereas at 170~K $I(Q,\omega=0)$ shows a detailed $Q$--dependence
arising from changes in $S(Q)$ (eq.~\ref{eq_two}).
The sharp maxima are Bragg--reflections of
crystalline cubic ice as it is presented in detail
in reference \cite{Geil-PCCP-2004}.

The temperature dependence of the elastic intensity $I(T,\omega=0)$
is reported in the right panel of Fig.~\ref{fig_04}.
Plotted data are taken at $Q=1.7$~\AA $^{-1}$ (Fig.~\ref{fig_04}~a)
and $Q=2.4$~\AA $^{-1}$ (Fig.~\ref{fig_04}~b).
Figure~\ref{fig_04}~c shows the signal integrated over
the accessed $Q$ range.
It is evident from Figure~\ref{fig_04}~c that despite
the intensity gain at the Bragg reflections (Fig.~\ref{fig_04}~a)
and intensity loss between (Fig.~\ref{fig_04}~b)
the total scattering power in the accessed $Q$ range
does not show any anomalies in the temperature range
of the supposed glass transition.
In conclusion, the behaviour of $I(T,\omega=0)$ in
Fig.~\ref{fig_04} gives evidence of a redistribution
of the signal within the elastic channel only,
which is compatible with a recrystallisation of the LDA matrix.
This recrystallisation is obviously slow at $T<150$~K but
very efficient above 150~K \cite{Hage-JCP-1994,Hage-JCP-1995},
causing the rapid gain in $I(\omega=0)$
shown in Fig.~\ref{fig_02}.
The very onset of the recrystallisation matches with the
supposed glass transition temperature $T_{\rm g}\approx 135$~K
that is indicated by the dashed vertical line in Figure~\ref{fig_04}.
\\

{\bf IV.b. Debye--Waller factor and mean--square displacement }\\

\begin{figure}
\begin{center}
\includegraphics[angle=0,width=140mm]{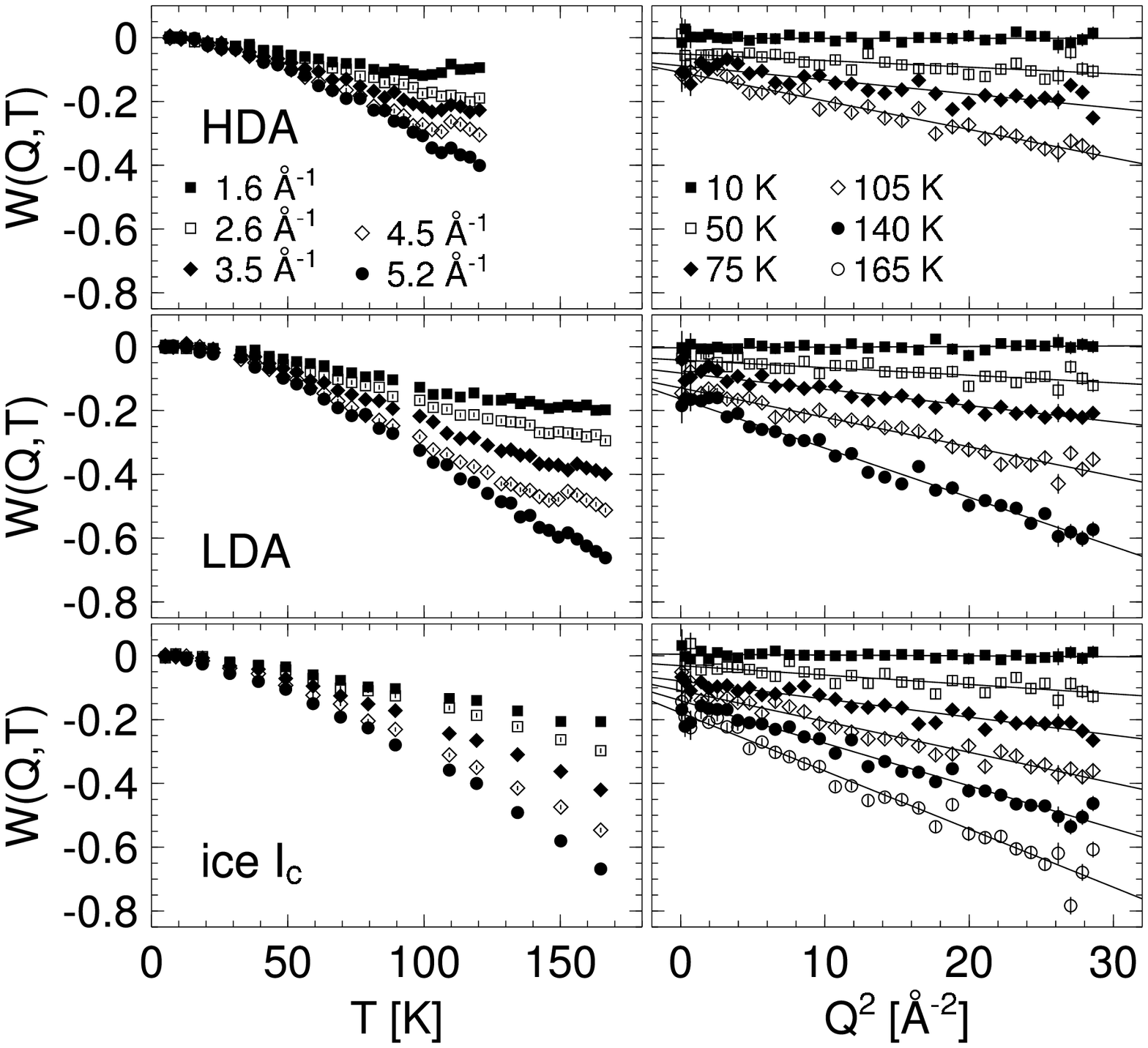}
\end{center}
\caption{Debye--Waller factor $W(Q,T)$ of HDA,
LDA and ice I$_{\rm c}$
measured with the partially protonated sample S1 on IN13.
Left panel gives the temperature dependence of $W(Q,T)$, right panel gives
the $Q$--dependence of $W(Q,T)$ plotted vs. $Q^2$ to stress
the behaviour expected for a harmonic solid.
The selected $Q$--values and approximate temperatures valid
for all presented data are given in the top subfigures.
Solid lines correspond to $Q^2$--fits to the data
whose slope is interpreted as the mean--square displacement
$\left< u^2(T)\right>$ in Fig.~\ref{fig_07}.}
\label{fig_05}
\end{figure}

\begin{figure}
\begin{center}
\includegraphics[angle=0,width=140mm]{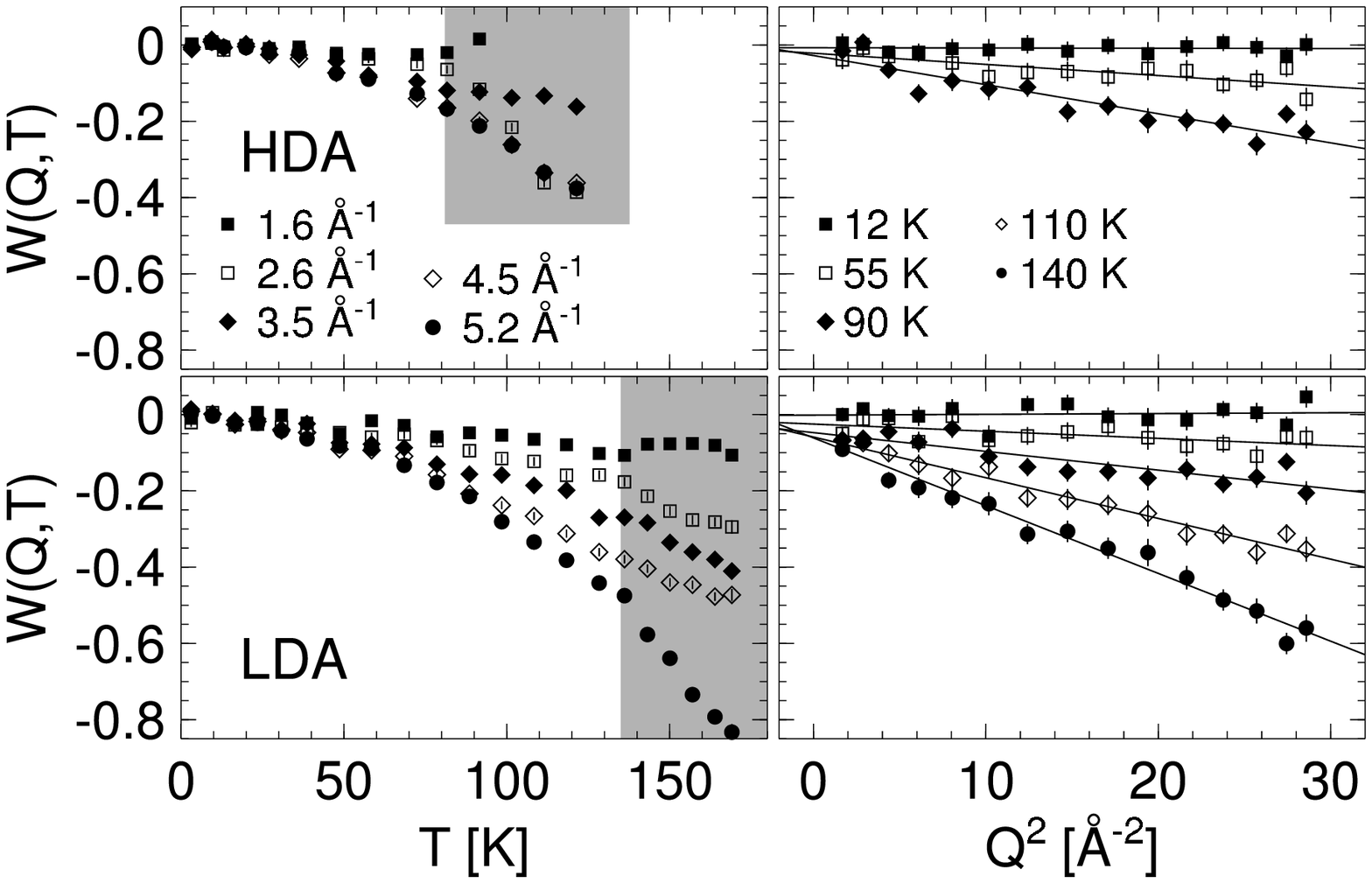}
\end{center}
\caption{Debye--Waller factor $W(Q,T)$ of HDA and
LDA measured with the deuterated sample S2 on IN13.
Left panel gives the temperature dependence of $W(Q,T)$,
right panel gives the $Q$--dependence of $W(Q,T)$ plotted
vs. $Q^2$ to stress the linear behaviour expected
for a harmonic solid.
The selected $Q$--values and approximate temperatures valid
for all shown data are given in the top subfigures.
Solid lines correspond to $Q^2$--fits to the data
whose slope is interpreted as the mean--square displacement
$\left< u^2(T)\right>$ in Fig.~\ref{fig_07}.
Please note that original data have been grouped by having
summed up two data points each.}
\label{fig_06}
\end{figure}

As it is outlined in section III
the Debye--Waller factor $W(Q,T)$ of a harmonic system
obeys the simple relation $W(Q,T)\propto - T \cdot Q^2$.
Figures~\ref{fig_05} and \ref{fig_06} show the $T$-- and
the $Q^2$--dependence of some selected $W(Q,T)$ data sets
for the samples S1 and S2 measured on IN13.
Since, the zero point oscillation of the scatterers is of no importance
for the discussion of $W(Q,T)$ here,
the presented data sets are normalised to the signals
measured at 2~K.
It is obvious from Fig.~\ref{fig_05} that all phases, HDA, LDA,
and ice I$_{\rm c}$, display a $Q^2$--dependence and are
(meeting the requirement of sufficiently high temperature)
proportional to $T$.
The linearity in $T$ is equally fullfilled for the elastic intensity
of S2 plotted in Fig.~\ref{fig_06} up to temperatures
close to the transformations of the phases.
Grey shaded areas in Fig.~\ref{fig_06} indicate the transformation
regions, in which the structural correlations change, as it is shown in
Fig.~\ref{fig_04}.
The $Q^2$--dependence is as well fulfilled for HDA and LDA
in S2 as it is for HDA, LDA and I$_{\rm c}$ in S1, which is demonstrated
by fits plotted as solid lines in Figs.~\ref{fig_05}
and \ref{fig_06}.

\begin{figure}
\begin{center}
\includegraphics[angle=0,width=110mm]{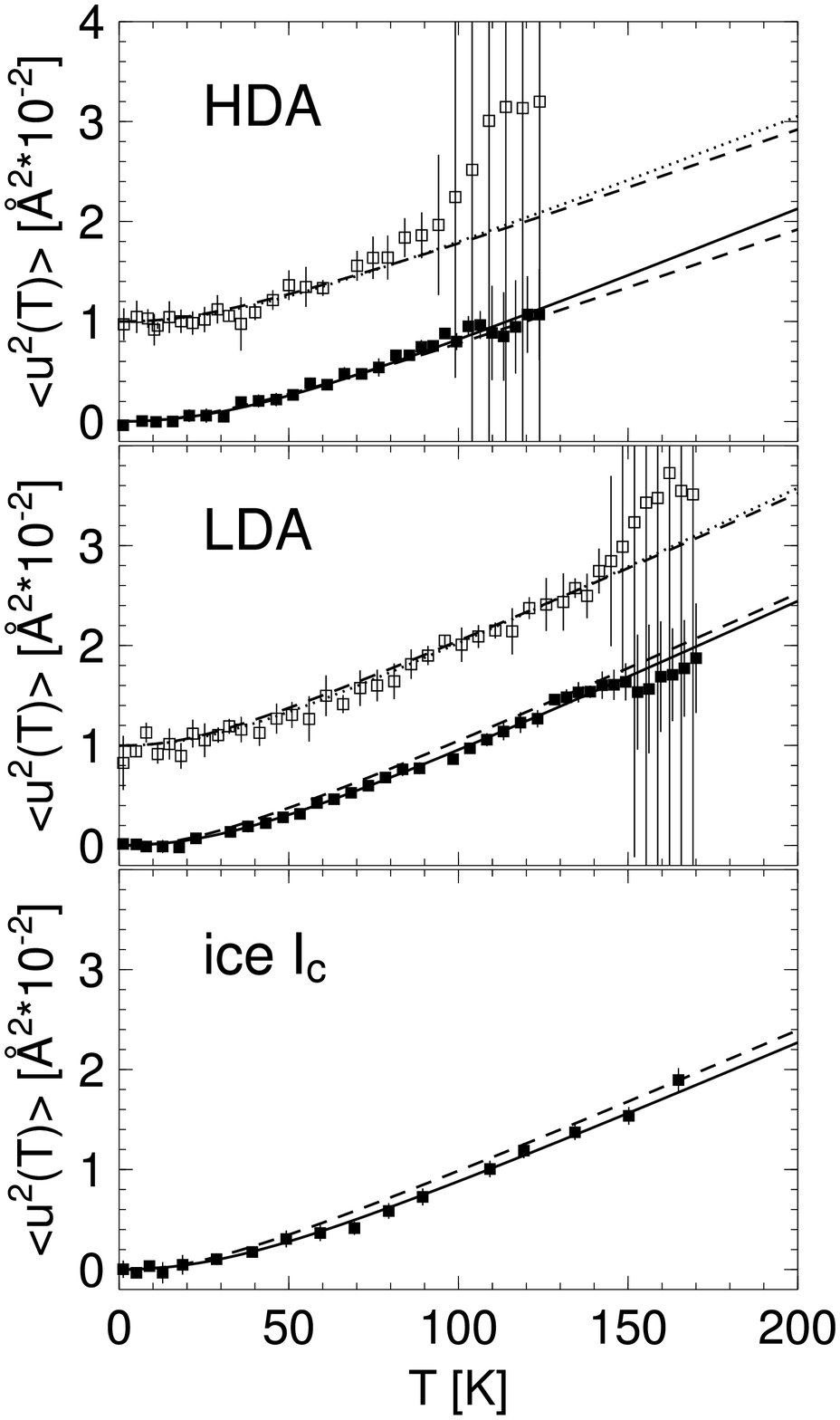}
\end{center}
\caption{Mean--square displacement $\left< u^2(T)\right>$
of HDA, LDA and ice~I$_{\rm c}$ determined from data
taken with sample S1 ($\blacksquare$) and with sample S2 ($\square$)
shown in Figs.~\ref{fig_05} and \ref{fig_06}, respectively.
Solid lines correspond to fits with the Debye--model to data taken
on S1, dotted lines correspond to fits to data on S2.
Dashed lines are calculated with eq.~\ref{eq_four} from the generalised
density of states given in Fig.~\ref{fig_08}.
Please note that data of sample S2 have been shifted for a better visibility.}
\label{fig_07}
\end{figure}

According to eq.~\ref{eq_three} the slopes of the fits correspond to the
mean--square displacement $\left< u^2(T)\right>$ of the scatterers.
They are shown for both S1 and S2 in Fig.~\ref{fig_07}.
Despite the different levels of deuteration of S1 and S2
the $\left< u^2(T)\right>$ of both samples in the HDA and LDA states
are comparable.
Please note that the onset of structural changes, which the samples
encounter during the phase transitions, is demonstarted
by the strong enhancement of statistical uncertaintity.
The comparable $\left< u^2(T)\right>$ values of both samples
in the corresponding phases can be understood when taking into
consideration that at the exploited temperatures $\left< u^2(T)\right>$
is strongly influenced by the mean velocity of sound (eq.~\ref{eq_nine}).
It has been shown that for the crystalline hexagonal ice I$_{\rm h}$
the level of deuteration does not change markedly the elastic constants
or thermodynamic observables being determined by the vibrational properties
\cite{Petrenko-Book,Ermolieff-SSC-1975,Leadbetter-PRS-1965}.
This means that the velocity of sound does not alter strongly
with the level of deuteration.
For example, a difference of only 3\% was found in specific heat
measurements of crystalline ice at $T\le 70$~K as an effect of deuteration
\cite{Leadbetter-PRS-1965}.
A similar behaviour can be expected in the phases studied here.

Assuming that the Debye model is applicable a Debye temperature $T_{\rm D}$
can be extracted from fits to $\left< u^2(T)\right>$
via eq.~\ref{eq_eleven}.
Fit results are plotted as solid (S1) and dotted (S2) lines
in Fig.~\ref{fig_07} and the obtained $T_{\rm D}$ are given
in Table~\ref{tab_two}.
The upper fitting limits are given by the transition
temperatures which are obtained with sample S2 and shown
in Fig.~\ref{fig_06}.

\begin{table}
\caption{\label{tab_two}
Debye--temperatures $T_{\rm D}$ obtained from fits to the
data shown in Fig.~\ref{fig_07} and estimated from velocity of sound
given in \cite{Gromnitskaya-PRB-2001}.  }
\begin{ruledtabular}
\begin{tabular}{cccc}
phase & sample S1 & sample S2 & from \cite{Gromnitskaya-PRB-2001} \\
\hline
HDA             & 230\,$\pm$\,6~K & 223\,$\pm$\,9~K & 230~K \\
LDA             & 217\,$\pm$\,2~K & 202\,$\pm$\,3~K & 203~K \\
ice I$_{\rm c}$ & 224\,$\pm$\,7~K & --    & 203~K \\
\end{tabular}
\end{ruledtabular}
\end{table}

Another observable in neutron scattering from which
$\left< u^2(T)\right>$ can be calculated is the density
of states $G(\omega)$ (eq.~\ref{eq_four}).
Figure~\ref{fig_08} reports $G(\omega)$ measured on the time--of--flight
spectrometer IN6
\cite{Koza-ILL-Report,Schober-PB-1998,Koza-PhD-2001}.
The plotted $G(\omega)$ are calculated from data measured
at 80~K (HDA), 127~K (LDA) and 160~K (ice I$_{\rm c}$).
Within the statistical accuracy of the measurements
performed at different $T$ no daviations from the presented
$G(\omega)$ could have been established.
A self consistent multi--phonon calculation was applied
to the data in addition to the standard correction procedures
\cite{Koza-PhD-2001,Reichardt-muphocor}.

\begin{figure}
\begin{center}
\includegraphics[angle=0,width=130mm]{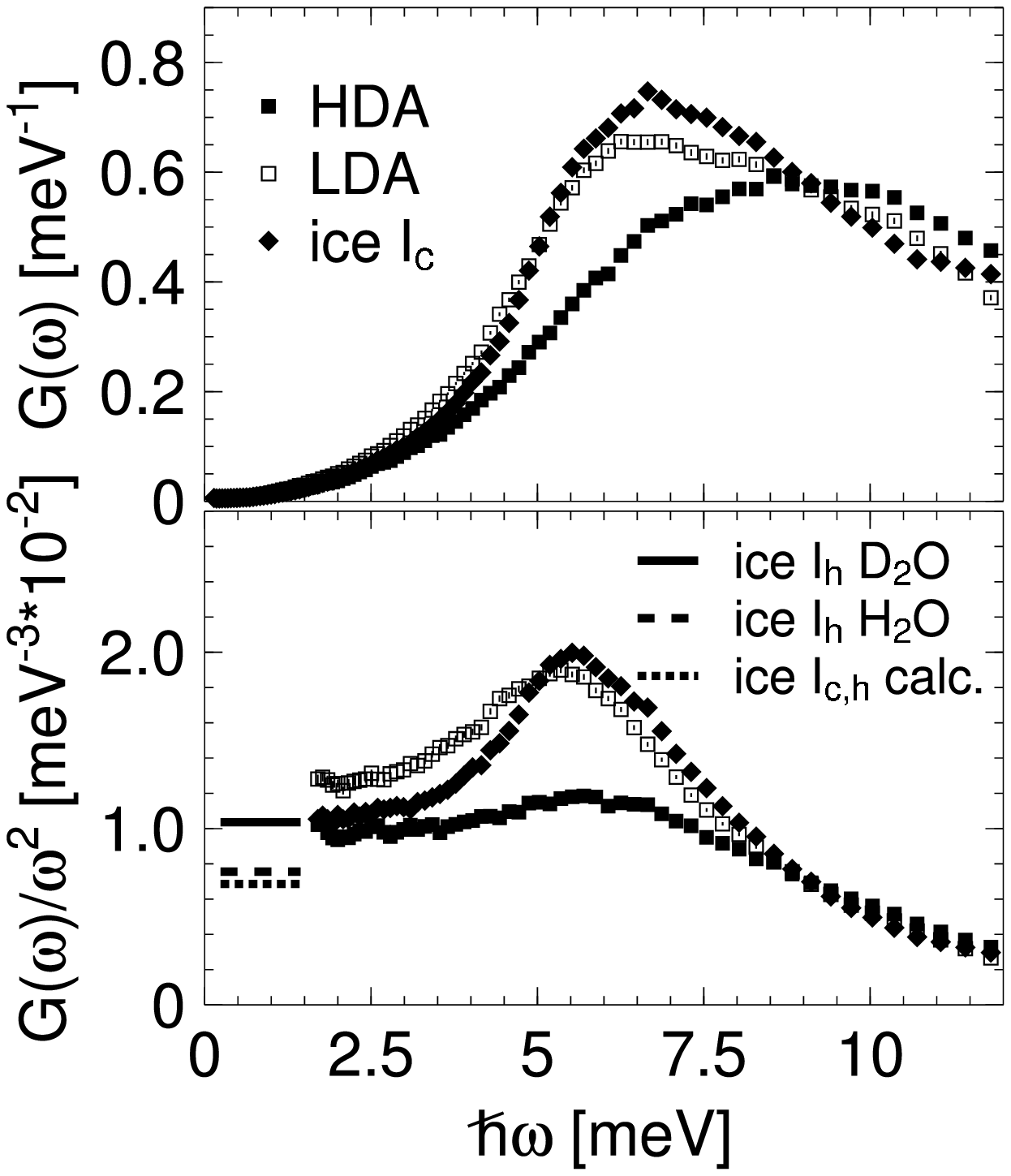}
\end{center}
\caption{Top: Generalised density of states $G(\omega)$
of HDA ($\blacksquare$), LDA ($\square$) and ice~I$_{\rm c}$
($\blacklozenge$) determined on the time--of--flight spectrometer
IN6 on a fully deuterated sample.
All $G(\omega)$ are normalised to 12 modes in the energy range
$\hbar\omega\le 40$~meV.
Bottom: $G(\omega)/\omega^2$ emphasises the low energy range
where a constant value is expected for harmonic solids.
Please note that data at $\hbar\omega\le 1.5$~meV are perturbed
by the resolution of the spectrometer and therefore suppressed
in the figure.
Equally shown are the Debye--levels determined independently
for fully deuterated (solid line) and fully protonated
(dashed line) ice I${\rm h}$, from which a factor of 0.7
accounting for the coherency effect is estimated.
Dotted line represents Debye--levels calculated from
\cite{Gromnitskaya-PRB-2001} for ice I$_{\rm c}$
and from \cite{Gagnon-JCP-1990} for ice I$_{\rm h}$
obtained by ultrasonic and light scattering experiments,
respectively.}
\label{fig_08}
\end{figure}

Since the data are obtained with a fully deuterated, coherently
scattering sample the determined $G(\omega)$ is an approximation
to the true density of states
\cite{Bredov-SPSS-1967,Oskotskii-SPSS-1967,Taraskin-PRB-1997}.
To account for the perturbations by the coherent scattering a
correction factor $\alpha=0.7$ is introduced.
This factor is estimated independently comparing $G(\omega)$
of hexagonal ice I$_{\rm h}$ and ice~XII
measured on fully deuterated and fully protonated
samples \cite{Koza-inprep}.
The Debye levels of the samples are indicated in
Fig.~\ref{fig_08} and compared to the Debye levels
obtained in ultrasonic experiments on ice I$_{\rm c}$
\cite{Gromnitskaya-PRB-2001} and Brillouin light scattering
on ice I$_{\rm h}$ \cite{Gagnon-JCP-1990},
both techniques giving comparable values.
Taking $\alpha$ and $G(\omega)$ from Fig.~\ref{fig_08} into account,
a $\left< u^2_{{\rm G}(\omega)}(T)\right>$ is calculated to the values
presented as dashed lines in Fig.~\ref{fig_07}.

Within the statistics of the measurements and the accuracy of the
applied data analysis the results obtained from backscattering
and time--of--flight measurements fully agree.
For example, in the framework of the Debye model
both techniques indicate the highest
average velocity of sound and, thus, the highest
$T_{\rm D}$ for HDA.
Debye temperatures estimated via eq.~\ref{eq_ten}
from results based on supersonic measurements \cite{Gromnitskaya-PRB-2001}
are included in Table~\ref{tab_two} and their respective
Debye levels for $G(\omega)$ are indicated in Fig.~\ref{fig_08}.
Although obtained with a different technique
the $T_{\rm D}$ match nicely the values observed in our elastic
neutron scattering experiments giving also a higher value
for $T_{\rm D}$ for HDA \cite{Comment-2}.

We would like to recall the fact that the coherent character of the sample introduces
uncertainties into  $G(\omega)$ although these are well controlled \cite{Koza-inprep}.
Thus, the excellent agreement of $\left< u_{{\rm G}(\omega)}^2(T)\right>$
with $\left< u^2(T)\right>$ calculated from the Debye--Waller
factor $W(Q,T)$ should not be over-interpreted.
However, qualitatively the results on $W(Q,T)$ and $\left< u^2(T)\right>$,
which are determined by studying the elastic signal,
and, on the other hand, the data on $\left< u_{{\rm G}(\omega)}^2(T)\right>$,
which is obtained from the inelastic response,
give a coherent picture of the properties
of HDA, LDA and ice I$_{\rm c}$.
They mark them unequivocally as harmonic systems.

Finally, we discuss the Debye temperatures determined
by specific heat measurements \cite{Handa-JCP-1986}
as 288~K for HDA, 305~K for LDA and 325~K for ice~I$_{\rm h}$.
The fundamental difference is that $T_{\rm D}$ is determined here
for $T \ll T_{\rm D}$, i.e. within the validity limits of the Debye model.
For this range specific heat measurements on ice~I$_{\rm h}$
give $T_{\rm D}$ values of about 200~K with a difference of
less than 3~\% at $T<70$~K between
deuterated and protonated samples \cite{Leadbetter-PRS-1965},
as it was already indicated above.
At elevated temparatures the specific heat reflects
the details of the phonon density of states $G(\omega)$ beyond
the acoustic region and the extracted
$T_{\rm D}$ becomes temperature dependent \cite{Ashkroft-Book}.
Taking only translational modes, i.e. phonons, into consideration
the $T_{\rm D}$ of ice~I$_{\rm h}$ reaches a maximum at $T\approx 100$~K
with $T_{\rm D}\approx 300$~K \cite{Leadbetter-PRS-1965}.
Consequently, the basic difference of about 100~K
between the $T_{\rm D}$ presented here and cited in \cite{Handa-JCP-1986}
is due to the applied experimental techniques and the thermodynamic
conditions of the measurements, but the data do agree in the general
concept of harmonic solids \cite{Ashkroft-Book,Lovesey-Book}.

The low value of $T_{\rm D}$ of HDA given in \cite{Handa-JCP-1986}
can be, on the one hand, understood by the temperature
sensitivity of $T_{\rm D}$ in the specific heat measurements.
At about 100~K not only phonons but also librations contribute
to $C_p(T)$ \cite{Leadbetter-PRS-1965,Koza-PhD-2001,Koza-inprep}.
It is well established by experiments that the librational band
shifts toward lower energies the higher the density of the
ice phase is.
The shift of the librational band in HDA raises the
specific heat and consequently pulls $T_{\rm D}$ below the value of
crystalline ice \cite{Schober-PB-1998,Klug-PRB-1991,Koza-PhD-2001}.
This effect is well comparable with the issue of deuterating
a crystalline sample \cite{Leadbetter-PRS-1965,Koza-PhD-2001}.
On the other hand, it must be stressed that the sample referred
to as HDA in \cite{Handa-JCP-1986} was annealed before the measurements
raising the question of the sample state.
In the case of a mixture of HDA--type and LDA--type phases
or even a contribution of ice~XII
\cite{Koza-JPCM-2003,Koza-inprep,Salzmann-PCCP-2003}
the density of states would decrease $T_{\rm D}$
more efficiently when increasing $T$.\\

{\bf VI.c. Excess of modes at low T}\\

So far we have not discussed the zero point oscillation
of the scatterers in HDA, LDA and ice I$_{\rm c}$
which is indicated in Fig.~\ref{fig_02}.
Indeed, bearing in mind that the elastic intensity is coupled to
the inelastic response (eqs.~\ref{eq_two}, \ref{eq_three},
\ref{eq_four}, \ref{eq_five})
there is an inconsistency concerning the behaviour of HDA.
Since, among all phases the lowest Debye--level is detected for
HDA (Fig.~\ref{fig_08}), it strictly requires the highest elastic
intensity $I(\omega=0)$ at $T\approx0$~K.
However, HDA exhibits the lowest $I(\omega=0)$ measured in all
our experiments (Fig.~\ref{fig_02}).
It must be notified that this inconsistency with
$I(\omega=0)$ holds also for results from other experiments
exploiting the dynamic response of the phases
\cite{Schober-PB-1998,Gromnitskaya-PRB-2001,Schober-PRL-2000,Koza-PRB-2004}.

Moreover, the Debye--Waller factors $W(Q,T)$, the mean--square
displacements $\left< u^2(T)\right>$, and the Debye temperatures
$T_{\rm D}$ calculated from the elastic scan measurements itself
display a dependence on temperature which is in full agreement
with the $G(\omega)$ determined by time--of--flight techniques.
Consequently, the temperature evolution of the elastic intensity
$I(\omega=0,T)$ of HDA contradicts its own value at very low
temperatures $I(\omega=0,T\rightarrow 0)$.

This self--contradicting behaviour of HDA can be resolved,
if not only phonons, i.e. excitations following the Bose
statistics, are inherent to the dynamic response but
also non--Bose modes predominant at low temperatures
are present.
Such modes have been detected for the first time
by three axis neutron scattering (TAS)
\cite{Svensson-PB-1994,Tulk-APA-2002},
confirmed later and interpreted as the contribution
of two--level systems (TLS) to the inelastic response of HDA
\cite{Agladze-PRL-1998,Agladze-EPL-2001,Agladze-PB-2002,Tse-PRL-2000}.
Remarkably, no TLS are observed in the LDA state.
Please note, that the level of deuteration of the sample has no
significant influence on the TLS intensity
\cite{Agladze-PRL-1998,Agladze-EPL-2001,comment-deuteration}.\\

{\bf V. SUMMARY AND CONCLUSIONS}\\

We utilised the neutron back--scattering and time--of--flight
techniques to gain information on the temperature dependence of
the high--density (HDA) and low--density (LDA) amorphous states
of ice and the crystalline phase I$_{\rm c}$.
Particular focus was put on the phase transformations
HDA~$\rightarrow$~LDA, LDA~$\rightarrow$~I$_{\rm c}$
and the supposed glass--transition of LDA with
$T_{\rm g}\approx 135$~K,
the mean--square displacements of the scatterers
in the stable states HDA, LDA and I$_{\rm c}$,
and their relative zero point oscillations.

Within the best energy resolution of 1~$\mu$eV of the experiments
no relaxation processes, giving evidence of a higher mobility
of the scatterers, can be identified during the phase
transformations.
From results on coherently scattering samples, we may conclude
that at or prior to $T\approx 135$~K, i.e., at the very $T_{\rm g}$
discussed in literature, the recrystallisation of LDA into
I$_{\rm c}$ sets in -- thus preventing any data analysis
in terms of Mode Coupling Theory as it is oulined in
\cite{Geil-PCCP-2004}.

The temperature and $Q$--dependence of the Debye--Waller factors
and the resulting mean--square displacements mark HDA, LDA and
ice I$_{\rm c}$ as harmonic solids.
In the framework of the Debye model HDA reveals a higher
Debye temperature, a higher average velocity of sound,
and thus a lower phonon density of states at low energy
when compared with LDA and ice I$_{\rm c}$.
LDA and ice I$_{\rm c}$ show rather similar dynamic properties.
The conclusions drawn from the elastic scan data on the temperature
dependence of all states are in full agreement with results
obtained by other experimental techniques.

The relative elastic intensities determined at low temperatures
support the existence of additional modes in HDA
not obeying the Bose statistics, i.e. this excess
should not be confused with a Boson--peak as it is identified
in glassy systems \cite{Sokolov-PRL-1997}.
These modes have been observed in inelastic neutron scattering
\cite{Svensson-PB-1994,Tulk-APA-2002},
confirmed and interpreted as two--level systems from light
absorption experiments
\cite{Agladze-PRL-1998,Agladze-EPL-2001,Agladze-PB-2002}.

In the case of the protonated samples, the scattering from
hydrogen is by a factor of $\sim$~40 stronger than the scattering
from oxygen.
Consequently, the signal measured with samples S3 and S4
is not only determined by the dynamic properties
of the water molecules but equally reflects the characteristics of
the hydrogen sublattice.
The absence of anomalies in the Debye--Waller factor excludes
therefore any strong proton relaxation within the matrix.
However, migration of lattice defects, as discussed in
references \cite{Fisher-JPC-1995,Fujara-JCP-1988,Kirschgen-PCCP-2003},
is not detectable due to their too low concentration.\\

{\bf ACKNOWLEDGEMENTS:}\\

We wish to thank D.D. Klug and N.I. Agladze for helpfull
discussions.


\begin{references}
\bibitem{Comment-0b}
                Please note that, although coherent and incoherent
                scattering has been exploited in this work
                we do not distinguish between the M\"ossbauer--Lamb
                and the Debye--Waller factor in the nomenclature.
                The last is used exclusively throughout the text.
\bibitem{Comment-0c}
                In the energy range accessed in the present experiments
                the dynamic properties of the samples are
                determined by vibrational modes of water molecules
                ($\hbar\omega < 40$~meV).
                The number of scatterers $N$ and their mass $M$ can be
                therefore related to the water molecule.
\bibitem{comment-iceIh}
               Please note, that the formation of ice~I$_{\rm h}$
               cannot be characterised by a specific transition temperature.
               The step at $T\approx240$~K in our data results
               from the thermal treatment, i.e. the heating rate,
               applied during the measurement. The formation
               of ice~I$_{\rm h}$ is stretched over a wide range of
               temperatures and was observed by us in other experiments
               on longer time scales already at about 200~K.
\bibitem{comment-deuteration}
It is noteworthy, that the TAS experiments
exploit the Stokes line, i.e. the neutron energy loss side,
whereas TOF experiments performed by our group
utilise the anti--Stokes side, i.e. the neutron energy gain side.
The difference between the TAS and TOF experiments is
manifested by the temperature dependence of the detailed balance factor
\cite{Lovesey-Book}, which results in a lower sensitivity
of the TOF experiments to the rather high energy
$\hbar\omega\approx 2$~meV of the TLS.
In the elastic mode, the back--scattering spectrometers IN13 and IN16
are sensitiv to both, the Stokes and the anti--Stokes line,
whereby the incident energy of neutrons determines the maximum
energy range on the Stokes side.
\end{references}
\end{document}